\documentclass[journal]{IEEEtran}
\usepackage[section] {placeins}
\usepackage{graphicx}
\usepackage{amsmath}
\usepackage{subfigure}
\usepackage{rotating}
\usepackage{multirow}
\usepackage{array}
\usepackage{rotating}
\usepackage{auto-pst-pdf}
\usepackage{graphicx}

\begin{document}

\title{REECH-ME: \underline{R}egional \underline{E}nergy \underline{E}fficient\\ \underline{C}luster \underline{H}eads based on \underline{M}aximum \underline{E}nergy\\ Routing Protocol for WSNs}

\author{A. Haider$^{1}$, N. Javaid$^{1,2}$, N. Amjad$^{1}$, A. A. Awan$^{1}$, A. Khan$^{3}$, N. Khan$^{3}$\\\vspace{0.4cm}
$^{1}$Dept of Electrical Engineering, COMSATS Institute of IT, Islamabad, Pakistan.\\
$^{2}$CAST, COMSATS Institute of IT, Islamabad, Pakistan.\\
$^{3}$Abasyn University, Peshawar, Pakistan.}
\maketitle

\begin{abstract}
\boldmath
In this paper, we propose Regional Energy Efficient Cluster Heads based on Maximum Energy (REECH-ME) Routing Protocol for Wireless Sensor Networks (WSNs) . The main purpose of this protocol is to improve the network lifetime and particularly the stability period of the network. In REECH-ME, the node with the maximum energy in a region becomes Cluster Head (CH) of that region for that particular round and the number of the cluster heads in each round remains the same. Our technique outperforms LEACH which uses probabilistic approach for the selection of CHs. We also implement the Uniform Random Distribution Model to find the packet drop to make this protocol more practical. We also calculate the confidence interval of all our results which helps us to visualize the possible deviation of our graphs from the mean value.

\end{abstract}

\section{Introduction}

A Wireless Sensor Network (WSN) can be reactive or proactive. In later case, nodes send their data to the Base Station (BS) or Cluster Head (CH) only when they detect a change and keep the transmitter off when they do not detect any change in the environment. Our proposed protocol is proactive. This approach is more energy efficient as compared to the reactive protocols. As in reactive protocols, nodes keep sending the data to the BS all the time. So, they quickly consume their energy as compared to the proactive protocols. In proposed protocol, the BS is at the centre of the field, i.e, if the area of the network is 100mx100m, the BS would be at a position (50m,50m) .

By the term homogenous, we mean that initially all nodes in the network have the same amount of energy. Similar to LEACH [1], REECH-ME is also based on the homogenous set of nodes. It all depends on the routing technique that how efficiently it consumes this energy to increase the life time and particularly the stability period of the network.

Clustering may be static or dynamic. In Static Clustering the clusters are not changed throughout the network life time. Whereas in Dynamic Routing, the clusters change depending on the network characteristics. LEACH uses Dynamic Clustering and its CHs are chosen on probabilistic basis. So the number of its CHs and the size of the clusters may change after every round. That is why its number of CHs is not optimum. So the number of packets sent to the BS is also not fixed as they depend upon the number of the CHs.

In the proposed scheme, the total area is divided into 9 regions. These are named as R1, R2, R3, ... , R9 as shown in Fig. 2. The region R1 is closest to the BS and uses Direct Communication as its routing technique. In Direct Communication, every node sends its data directly to the BS. All other regions, i.e R2 - R9, do not use Direct Communication. Instead, they form CHs to send their data to the BS. REECH-ME uses Static Clustering, so clusters throughout the network lifetime remain the same. Each region except R1 is called a cluster and each cluster has only one CH for a particular round. Other nodes of regions R2-R9 send their data to the BS via CH of their region. In our protocol, the CH is chosen on the basis of maximum energy. It means that in any round the node having the maximum energy becomes the CH. So the energy utilization becomes very efficient as well as the number of the CHs in a round becomes fixed. As there are 8 regions which form clusters, so there would be 8 CHs in each round which is the optimum number.

As in any real case scenario, the number of packets received at the BS is never equal to the number of packets sent to the BS. This is because some packets are lost due to certain factors. Those factors may include interference, attenuation, noise, etc. That is why we use the Uniform Random Distribution Model [5] for the calculation of packets drop. This makes REECH-ME more practical.

We also calculate the Confidence Interval of all our results. It helps us to visualize the deviation of the graphs from the mean value. Where, the mean value is calculated by taking the results of 5 simulations, and then taking their mean.

\section{Motivation}
The main objective of a routing protocol is to efficiently utilize the energy of the nodes. This is because these nodes are not rechargeable and in order to make them useful for a longer period of time, routing protocols have been proposed. Routing protocols improve the lifetime of a network and specifically the stability period of a network. Protocols [1] , [6] , [7], [11], [12], [13], [17], [18], [19] and [20] are proposed to achieve these goals. As shown in Figure~1, LEACH uses dynamic clustering. Hence, its clusters change after every round.

\begin{figure}[!h]
\centering
\includegraphics[height=8cm, width=8cm]{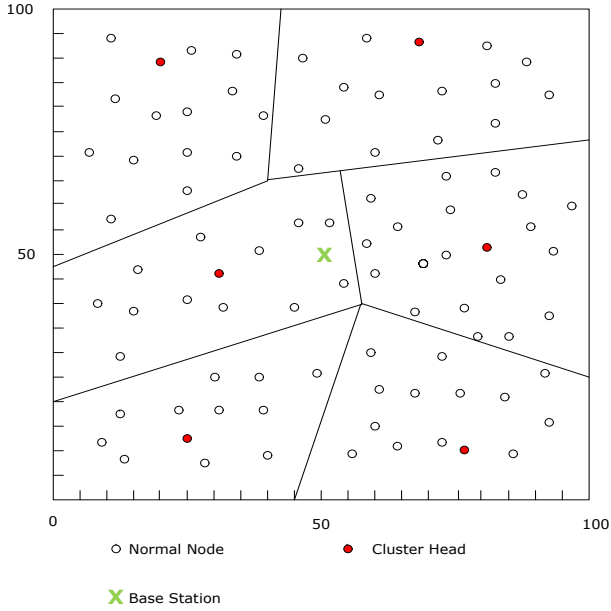}
\caption{Clustering in LEACH Protocol}
\end{figure}

As the CH selection in LEACH is on the basis of probability, the optimum number of CHs is not achieved. So the energy is not efficiently utilized. The area coverage in LEACH is also not very efficient. This is because it treats the whole area as a single area and the nodes are deployed in it at once. So some of the area is left unattended. To efficiently utilize the energy and to improve the coverage area, many researchers have introduced some effective approaches [2] , [3] , [4] and [10]. In these approaches, the total area is divided into small regions and these regions are treated separately for the nodes distribution and it improves the area coverage. In our protocol, we also use the approach of dividing the total area into smaller areas. We use the direct transmission for the area (R1) closest to the nodes as shown in Fig. 2. We use the static clustering in all other regions. The CH selection is based on the maximum energy of a particular node in a round. It means that the node with the highest energy is chosen as the CH for that particular round. So the energy is very efficiently utilized and the area coverage is also improved.

\section{Radio Model}
  REECH-ME assumes a simple first order radio model in which the radio dissipates $E_{elec}$ = 50 nJ/bit for powering the transmitter or receiver circuitry and $E_{amp}$ = 100 pJ/bit/$m^2$ for the transmit amplifier to achieve an acceptable Eb/No. Transmitter circuitry also consumes $E_{DA}$ = 50 nJ/bit to aggregate the data received by the normal nodes. We also take in account the $d^2$ energy loss due to channel transmission. Thus, to transmit a k-bit message distance d the energy is given as:

\begin{equation}
  d_{o}=\sqrt\frac{\epsilon_{fs}}{\epsilon_{mp}}
\end{equation}
if $d<d_{o}$
\begin{equation}
  E_{Tx}(k,d)=E_{elec} * k + \epsilon_{fs}*k*d^2
\end{equation}
if $d \geq d_{o}$
\begin{equation}
  E_{Tx}(k,d)=E_{elec} * k + \epsilon_{mp}*k*d^4
\end{equation}
 Reception Energy:
\begin{equation}
  E_{Rx}(k)=E_{elec} * k
\end{equation}

Where $E_{elec}$ is the energy dissipated per bit to run the transmitter or receiver circuit, $\epsilon_{fs}$ and $\epsilon_{mp}$ depend on the transmitter amplifier.

\section{The REECH-ME Protocol}
An efficient routing protocol is the one which consumes minimum energy and provides good coverage area. Minimum consumption of energy leads towards better network lifetime and particularly the stability period. Whereas good coverage area is useful in getting the required information from the whole network area. Because if the coverage area is not good, then their would be some small areas left unattended in the network. These unattended areas are referred to as coverage hole. The primary objective of a routing protocol is to achieve minimum energy utilization and full coverage area. Many researches have addressed such matters as in [2] and [3]. Different approaches are used to solve this problem, one of which was the division of the network field area into sub areas. In the proposed technique, we divide the network area into sub areas as explained in the following subsection.

\subsection{Formation of Regions}
In LEACH, the CHs are elected on probabilistic basis and threshold is calculated for each node. Cluster is formed on the basis of received signal strength from the CH and its associate nodes. In our protocol, we divide the area in different regions as shown in Fig. 1. First of all, the whole area is divided into two concentric squares. The inner square is itself a region and is referred to as Region 1 or R1. The outer square is divided into 8 regions, 4 of which are rectangles and 4 are squares as shown in Fig. 2. The boundaries of all regions are taken as:

\begin{itemize}

  \item R1 - (25 - 75, 25 - 75)

  \item R2 - (50 - 100, 75 - 100)

  \item R3 - (0 - 25, 75 - 100)

  \item R4 - (0 - 25, 50 - 75)

  \item R5 - (0 - 25, 25 - 50)

  \item R6 - (0 - 50, 0 - 25)

  \item R7 - (50 - 100, 0 - 25)

  \item R8 - (75 - 100, 25 - 50)

  \item R9 - (75 - 100, 50 - 75)
\end{itemize}

Each region contains fixed number of nodes. R1 contains 20 nodes, whereas, regions R2-R9 contain 10 nodes each. The BS is located at the center of the field. Fixed number of nodes are randomly distributed in their defined regions.

\begin{figure}[!h]
\centering
\includegraphics[height=8cm, width=8cm]{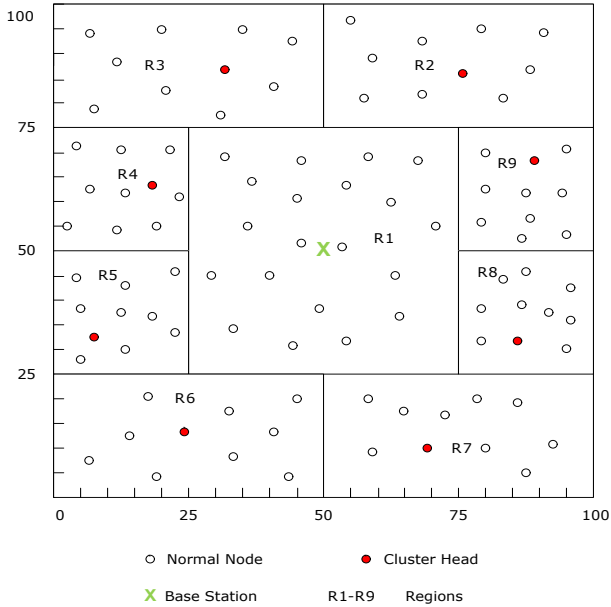}
\caption{Regions in REECH-ME}
\end{figure}

\subsection{CH Selection}
Unlike LEACH in which the CHs are selected on probabilistic basis, REECH-ME selects a node as the CH of that region if it has the maximum energy before the start of that round. Initially, all nodes have the same amount of energy and any node can become the CH for first round. So, a node is chosen randomly to become the CH of that region for the first round. All other nodes send their data to CH which receives the data from all the nodes, aggregates it and sends it to the BS. When the first round is completed, the amount of energy in each node would not be the same. This is because the utilization of energy depends upon the distance between the node/CH which is transmitting and the CH/sink which is receiving. The larger the distance, the greater energy is consumed. And smaller the distance, smaller energy is consumed. As distance for transmission and reception is different for different nodes, the energy consumption will also be different for different nodes. For every next round, the CH is selected on the basis of their energies. The node with the maximum energy in a region becomes the CH of that region for that particular round. All the regions except R1 will follow the same technique of CH selection.

\section{Simulations and Results}
In this section, we assess the performance of our protocol using MATLAB. In our protocol total area is divided into 9 regions. Region 1 uses direct communication as its routing technique. Whereas, all other regions use clustering which is based on maximum energy of a node in that particular region. The node with the maximum energy in a particular region becomes the CH of that region. Normal nodes of a region send their sensed data to BS via CH of their own region. In this way, after every round, a new node which has the maximum energy in that region is chosen as the CH of its region. The simulation parameters are given in Table 1.

\begin{center}
\begin{tabular}{p{5cm} l  l}
  Table 1&~\\
  Parameters used in Simulations&~\\
  \hline
  Parameter & Value \\
  \hline
  Network Size & 100m x 100m \\ 
  Node Number & 100 \\ 
  Initial Energy of Normal Nodes & 0.5J \\ 
  $E_{TX}$ & 50nJ \\ 
  $E_{RX}$ & 50nJ \\ 
  $E_{DA}$ & 5nJ \\ 
  Packet Size & 4000 bits\\
  Probability of Packet Drop & 0.3\\
  Sink Location & (50m,50m)\\
  \hline
  ~&~\\
\end{tabular}
\end{center}

\subsection{Performance Parameters}

In the following subsections of performance parameters, we will discuss confidence interval, network lifetime, throughput and packet drop.
\subsubsection{Confidence Interval}
The nodes are randomly distributed in a certain region. They may be placed any where in a particular region. Any new distribution change the location of nodes in network area. In this way, the calculations regarding their lifetime, stability, instability region, packet drop, etc. slightly vary. So, keeping this fact in mind, we also calculated the confidence interval of all our results. Confidence interval helps us to visualize the deviation of the graphs from the mean value. Where, the mean value is calculated by carrying out the simulations for 5 times, and then taking their mean. We calculate the confidence interval of all our graphs.

\subsubsection{Network Lifetime}
Alive nodes refer to those nodes which have sufficient energy to sense and transmit data. The lifetime of a network depends upon the number of alive nodes. As long as there is even one alive node in the network, its lifetime counts. So the lifetime of a network refers to the time period from the start of the network till the death of the last node. First of all, we compare the lifetime of LEACH with our REECH-ME. The Fig. 3 shows the confidence interval of dead and alive nodes. We calculate the confidence interval because it helps us to visualize the deviation of the graph from its mean value. Whereas, the mean value is calculated by carrying out 5 simulations and then taking their mean.

\begin{figure}[!h]
\centering
\includegraphics[height=6cm, width=8cm]{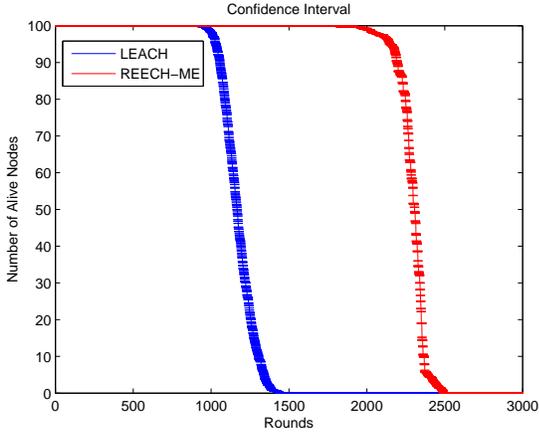}
\caption{Number of Alive Nodes}
\end{figure}

Fig. 3 shows the number of alive nodes. It can be seen that the network lifetime of our protocol is 66\% more than that of the LEACH, i.e, around 2500 and 1500 rounds respectively. The stability period is a time duration from the start till the death of the first node. The stability period of our protocol is 79\% better than the LEACH.

REECH-ME uses maximum energy based CH selection. Whereas in LEACH, the clustering is based on the probability. Maximum energy based clustering helps to utilize the energy of only those nodes which have the maximum energy in their regions. So the energy of all nodes is very efficiently utilized.

We always obtain the optimum number of CHs in a round, i.e 8 because we divide the whole area into 9 smaller regions. And 8 regions use clustering and each region has only one CH. So the number of clusters and CHs is always fixed. Whereas in LEACH, the number of CHs is never the same and hence, the energy utilization is not efficient. The instability period is the time duration between the death instants of the first node and the last node alive in the network. The instability region in our protocol is 40\% more than that of LEACH.
\begin{figure}[!h]
\centering
\includegraphics[height=6cm, width=8cm]{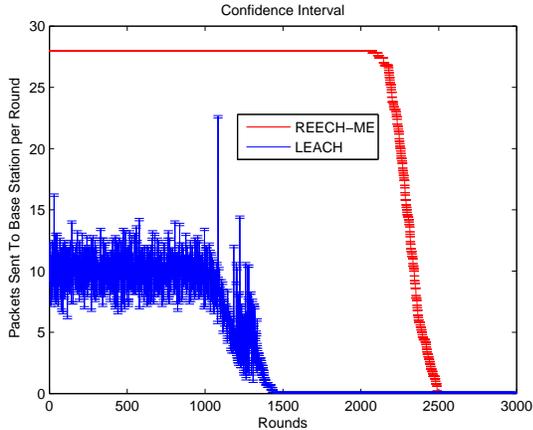}
\caption{Number of Packets Sent to BS Per Round}
\end{figure}

\subsubsection{Packets Sent to BS}
The average packets sent to the sink in LEACH are less as compared to REECH-ME as shown in Fig. 4. This is because on an average, there would be around 10 CHs (not always exactly 10) in a round. And we know that the normal nodes do not send their data directly to the sink. Instead, they send their data to the BS via the CH. So on an average, there would be around 10 packets sent per round. Whereas in our Protocol, 20 nodes are present in the region which is closest to the sink and they send their data directly to the sink. In all the other 8 regions, 8 nodes would be CHs in each round. So, on an average there would be 28 packets sent per round. As the number of the dead nodes increases, the number of packets sent to the BS decreases. In LEACH, the first node dies in approximately 1000 rounds. So, after that round, the number of packets sent to the sink also gradually decreases in correspondence with the number of dead nodes. Similarly, in our Protocol, the average number of packets received also gradually decreases with the increase in number of dead nodes.

\subsubsection{Packet Drop}
Ideally when a CH sends its data to the BS, all the packets are received successfully without any loss, i.e, the number of packets sent to the BS are equal to the number of packets received at the BS. But in reality it does not happen. Whenever the data is sent to BS from a CH, some of its packets do not reach the destination. This is called Packet Drop. The reason behind this packet drop may be the interference, attenuation, noise, etc. In our protocol, we have implemented the uniform random distribution to calculated the packet drop. This makes our protocol more practical. We used 0.3 as the packet drop probability value. Fig. 5 shows the number of packets sent to the BS per round, whereas, Fig. 4 shows the number of packets received at the BS. Its can be observed that the number of packets received at the BS is less than the number of packets sent to the BS. Thus, packet drop makes our protocol more applicable and practical as well.

\begin{figure}[!h]
\centering
\includegraphics[height=6cm, width=8cm]{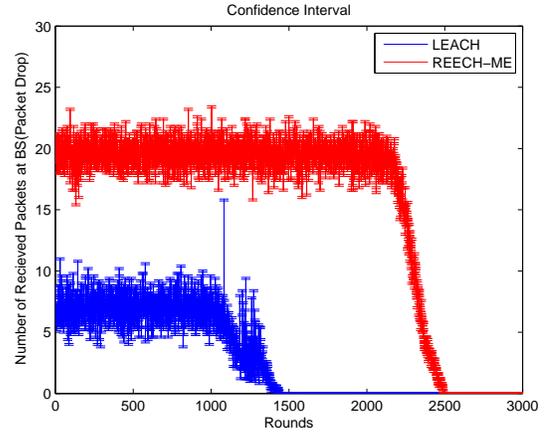}
\caption{Number of Packets Received at Sink after Packet Drop}
\end{figure}


Ideally, whenever the data is sent to the sink, it reaches without any packet loss. But in real situations this ideal condition does not exist. That means that when data is sent to the sink from the CHs, some packets are lost. To show this packet loss, we use the Uniform Random Distribution Model to find the packet drops. We calculate the packet drop by taking the packet drop probability as 0.3 and then calculate the confidence interval as shown in Fig. 5. Due to the packets drop, the number of packets that is received at the sink would be less as compared to the number of packets sent by the CHs. So in our protocol, the number of packets received at the BS fluctuates around 20 packets. Whereas in LEACH, the number of packets received at the BS fluctuates around 7. The number of received packets decreases as the number of dead nodes increases. As the stability region of LEACH is smaller as compared to our protocol, the number of received packets starts to decrease from around 1000th round. Whereas in our protocol, this decrement in the number of the received packets starts from around 1800th round.

In REECH-ME the average number of packets that is sent by the nodes to the BS is 28. By applying packet drop probability of 0.3, the average number of packets which are dropped is around 8. Similarly on an average the total number of packets which are sent to the BS from the nodes is 10. And 0.3 probability of packet drop does not allow all the packets to reach the destination , i.e. BS. The average number of packets dropped due to this probability in LEACH is 3. This behavior can be seen in Figure~6. The number of packets dropped in REECH-ME is more as compared to the number of packets dropped in LEACH. And the reason behind it is that in REECH-ME the average number of packets which is sent to the BS is more than that of the LEACH. The 0.3 probability on both the protocols will result in different number of packets dropped in both protocols.

\begin{figure}[!h]
\centering
\includegraphics[height=6cm, width=8cm]{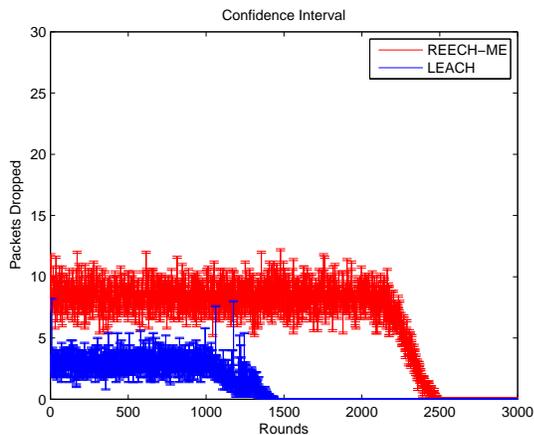}
\caption{Number of Packets Packets Dropped}
\end{figure}

\section{Conclusion and Future Work}
Our proposed technique uses static clustering and CHs are selected on the basis of the maximum energy of the nodes. This results in fixed number of CHs in each round and the optimum number of CHs is also maintained. We implemented Packet Drop Model to make our protocol more practical. We also implemented confidence interval to find the possible deviation of our graphs from the mean value, where mean value is calculated by simulating our protocol 5 times and then taking its mean. We compare the results of our protocol with that of the LEACH. REECH-ME outperforms LEACH in network lifetime, stability period, area coverage and throughput. Thus, this scheme enhances the desired attributes, i.e, minimum energy consumption, maximum stability period, better lifetime and throughput allot as compared with LEACH.

In future, Routing Link Matrices can also be applied on this proposed technique. Routing can be done by adapting many different approaches as done in [14], [15] and [16]. Application of Routing Link Matrices on the proposed scheme can be useful in achieving efficient consumption of energy in the network.

\end{document}